\documentclass{webofc}
\usepackage[varg]{txfonts}   % Web of Conferences font
\usepackage[T1]{fontenc}
\usepackage{geometry}
\usepackage{color}
\usepackage{amsbsy}
\usepackage{amssymb}
\usepackage{graphicx}
\usepackage{setspace}
\begin{document}
\title{ Resonance production in high energy collisions from small to big systems}
\author{{
K.$\,$Werner$^{(a)}$, A. G. Knospe$^{(b)}$, C. Markert$^{(c)}$, B. Guiot$^{(a,d)}$, Iu.$\,$Karpenko$^{(a,e,f)}$,
T.$\,$Pierog$^{(g)}$ , G. Sophys$^{(a)}$, M. Stefaniak$^{(a,h)}$,
M. Bleicher$^{(i,j)}$, J. Steinheimer$^{(i)}$\\
\\
$^{(a)}$ SUBATECH, University of Nantes -- IN2P3/CNRS -- IMT Atlantique, Nantes, France\\
$^{(b)}$ University of Houston, Houston, USA\\
$^{(c)}$ University of Texas, Austin, USA\\
$^{(d)}$ Universidad Tecnica Federico Santa Maria y Centro Cientifico-\\
\hspace*{0.5cm} Tecnologico de Valparaiso, Valparaiso, Chile\\
$^{(e)}$ Bogolyubov Institute for Theoretical Physics, Kiev 143, 03680, Ukraine\\
$^{(f)}$ INFN - Sezione di Firenze, Via G. Sansone\,1, I-50019 Sesto Fiorentino (Firenze), Italy\\
$^{(g)}$ Karlsruhe Inst. of Technology, KIT, Campus North, Inst. f. Kernphysik, Germany\\
$^{(h)}$ Warsaw University of Technology, Warsaw, Poland\\
$^{(i)}$ FIAS, Johann Wolfgang Goethe Universitaet, Frankfurt am Main, Germany\\
$^{(j)}$ GSI Helmholtzzentrum, Planckstr. 1, 64291 Darmstadt\\
}
}

\abstract{%
The aim of this paper is to understand resonance production (and more
generally particle production) for different collision systems, namely
proton-proton (pp), proton-nucleus (pA), and nucleus-nucleus (AA)
scattering at the LHC. We will investigate in particular particle
yields and ratios versus multiplicity, using the same multiplicity
definition for the three different systems, in order to analyse in
a compact way the evolution of particle production with the system
size and the origin of a very different system size dependence of
the different particles.
}
\maketitle

The main goal of heavy ion physics at very high energies is the proof
of existence of the creation of a quark-gluon plasma, and the study
of the properties of this exotic state, by analysing the final state
of many thousands of produced hadrons. Unfortunately, many of these
final state particles are not directly coming from the decay of the
plasma, but they are produced or at least affected by the hadronic
cascade, the last stage of the collision before particles freeze out
and freely move to the detectors. Here resonance studies come into
play : There is a large number of resonances at our disposal, having
very different lifetimes, from 1 fm/c to several tens of fm/c, which
means that these particles decay with varying probabilities in the
hadronic stage, and provide therefore valuable information about the
latter one.

We extended this analysis, to also consider small collision systems
(pp and pA), in addition to AA. We also consider not only resonances,
but also stable particles, providing useful additional information
(the lifetime is not the only relevant parameter).

Our tool to analyse particle production is the EPOS model. EPOS3 \cite{epos3}
is a universal model in the sense that for pp, pA, and AA collisions,
the same procedure applies, based on several elements:
\begin{description}
\item [{Initial~state.}] A Gribov-Regge multiple scattering approach is
employed (``Parton-Based Gribov-Regge Theory'' PBGRT \cite{hajo}),
where the elementary object (by definition called Pomeron) is a DGLAP
parton ladder. Starting from a multi-Pomeron structure for the elastic
scattering S-matrix, one then uses cutting rules to obtain (partial)
cross sections of inelastic processes, by employing Markov chain techniques.
\begin{figure}[tb]
\noindent \begin{centering}
\hspace*{1cm}(a)%
\begin{minipage}[c]{0.55\columnwidth}%
\noindent \includegraphics[scale=0.34]{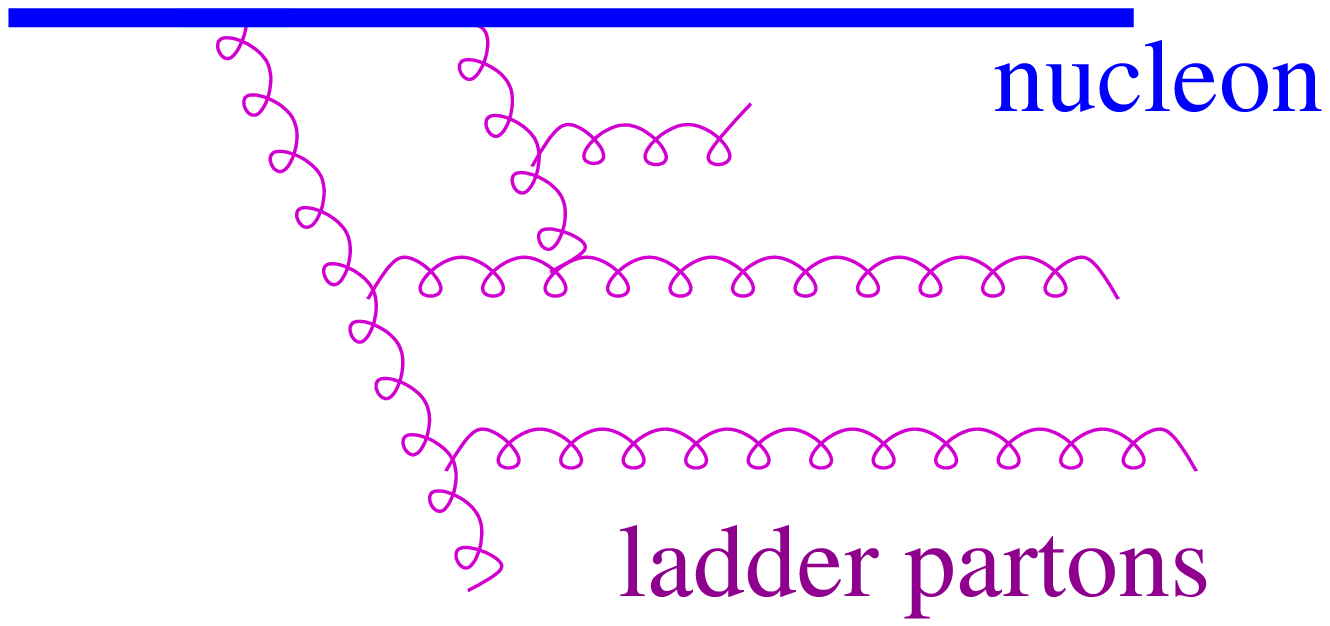}%
\end{minipage}(b)%
\begin{minipage}[c]{0.4\columnwidth}%
\noindent ~~~~~~~\includegraphics[scale=0.35]{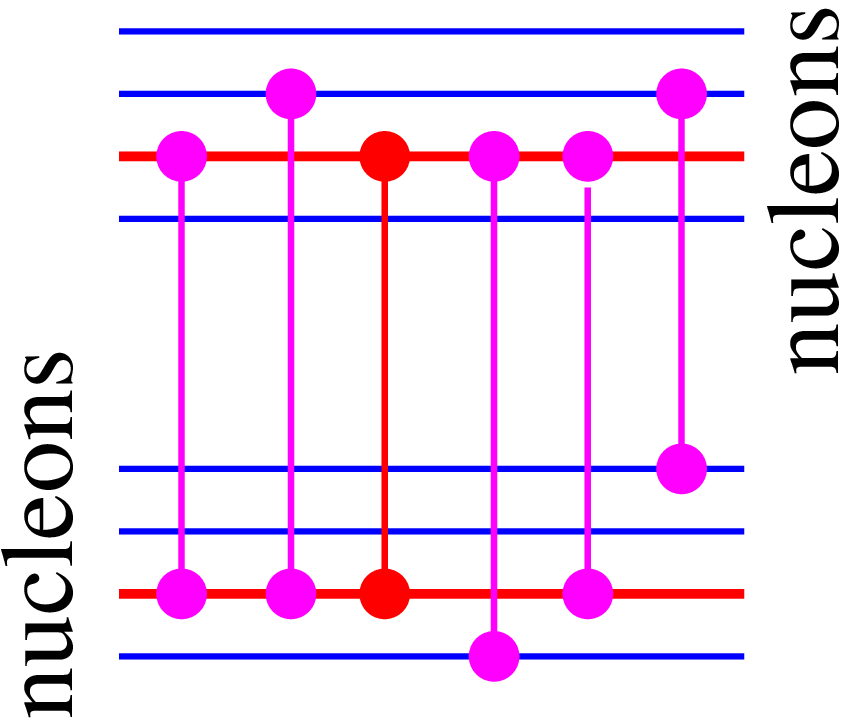}%
\end{minipage}
\par\end{centering}

\protect\caption{(Color online) (a) Non-linear effect: ladder fusion. (b) Pomerons
connected to a given (red) Pomeron \label{fig-1}}
 
\end{figure}

\item [{Non-linear~effects.}] High parton density effects like gluon fusion
during the parton evolution (see fig. \ref{fig-1}(a)) are taken into
account by using a dynamical saturation scale for each Pomeron, of
the form\textbf{ $Q_{s}=Q_{s}(N_{\mathrm{Pom}},\,\hat{s})$}, depending
on the number of Pomerons $N_{\mathrm{Pom}}$ connected the Pomeron
in question (see fig. \ref{fig-1}(b)), and its energy $\hat{s}$.
The functional form of $Q_{s}$ is obtained from fitting to the energy
dependence of basic cross sections and the multiplicity dependence
of particle production in pp at very high transverse momenta.
\item [{Core-corona~approach.}] The parton ladders corresponding to Pomerons
are treated as classical relativistic (kinky) strings. So in general,
we have a large number of (partly overlapping) strings. Based on the
momenta and the density of string segments, one separates at some
early proper time $\tau_{0}$ the core (going to be treated as fluid)
from the corona (escaping hadrons, including jet hadrons). The core-corona
procedure has been first described in \cite{core}, a more recent
discussion is found in \cite{epos3}. The corresponding energy-momentum
tensor of the core part is transformed into an equilibrium one, needed
to start the hydrodynamical evolution. This is based on the hypothesis
that equilibration happens rapidly and affects essentially the space
components of the energy-momentum tensor.
\item [{Viscous~hydrodynamic~expansion.}] Starting from the initial proper
time $\tau_{0}$, the core part of the system evolves according to
the equations of relativistic viscous hydrodynamics \cite{epos3,yuri},
where we use presently $\eta/s=0.08$. A cross-over equation-of-state
is used, compatible with lattice QCD \cite{lattice,kw1}. The ``core-matter''
hadronizes on some hyper-surface defined by a constant temperature
$T_{H}$, where a so-called Cooper-Frye procedure is employed, using
equilibrium hadron distributions, see \cite{kw1}.
\item [{Final~state~hadronic~cascade.}] After hadronization, there occur
still hadron-hadron rescatterings, realized via UrQMD \cite{urqmd}.
\end{description}
The above procedure is employed for each event (event-by-event procedure).\\

We will discuss particle production as a function of the average multiplicity
per eta interval $\left\langle dn/d\eta(0)\right\rangle $ at central
pseudorapidity ($\eta=0$), whereas forward pseudorapidity intervals
are used to define the different multiplicity classes. This is the
same procedure as used by the ALICE experiment. In the following $\left\langle dn/d\eta(0)\right\rangle $
is simply referred to as ``multiplicity''. 

\begin{figure}[tb]
\noindent \begin{centering}
\begin{minipage}[c]{0.33\columnwidth}%
\noindent \begin{center}
\textbf{\textcolor{red}{central AA}}\\
\includegraphics[scale=0.22]{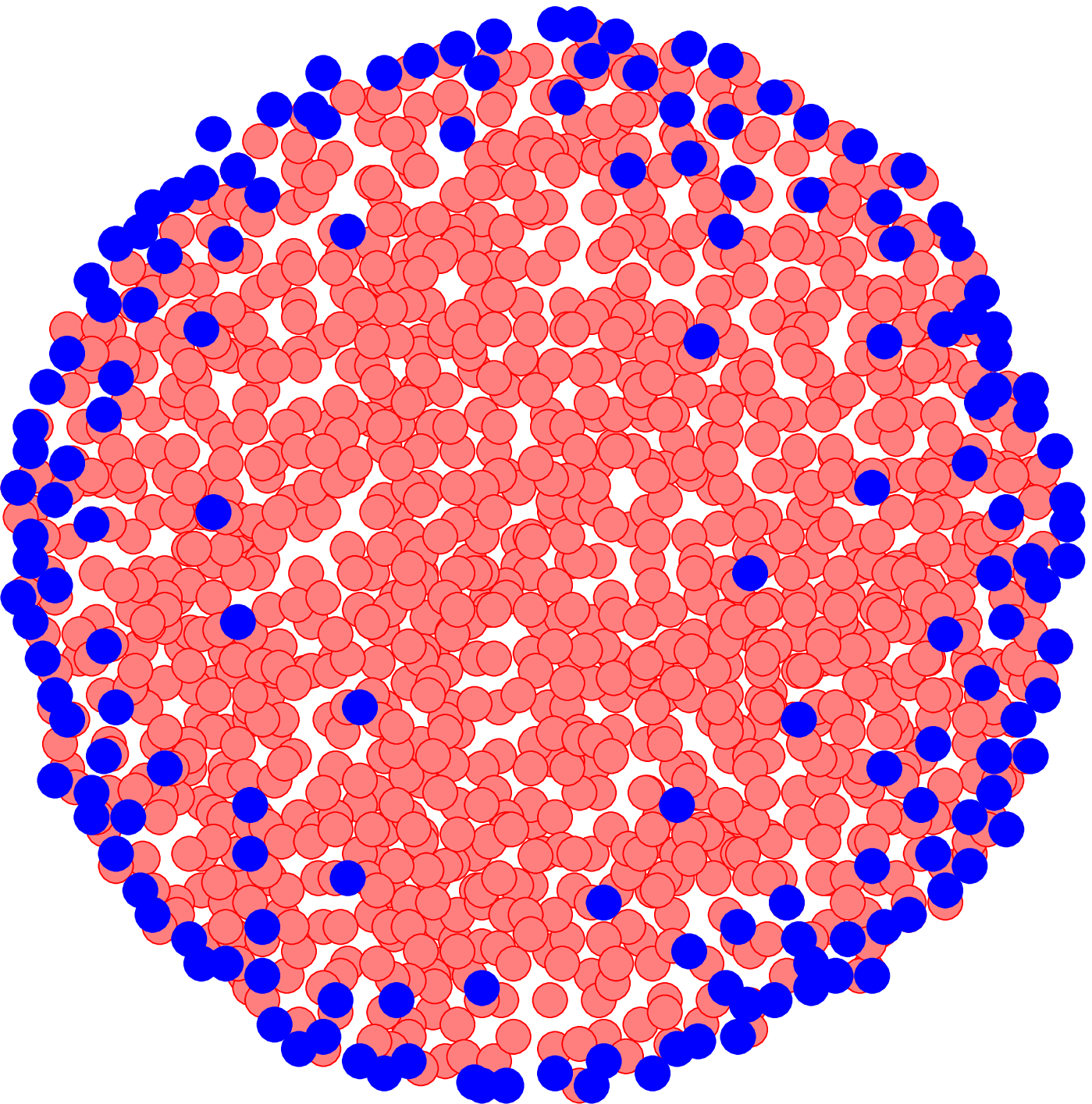}
\par\end{center}%
\end{minipage}%
\begin{minipage}[c]{0.33\columnwidth}%
\noindent \begin{center}
\textbf{\textcolor{red}{peripheral AA}}\\
\textbf{\textcolor{red}{high mult pp,pA}}\\
\includegraphics[scale=0.22]{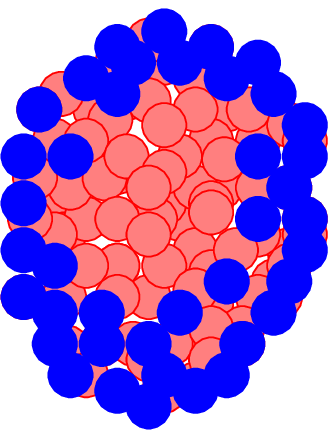}
\par\end{center}%
\end{minipage}%
\begin{minipage}[c]{0.33\columnwidth}%
\noindent \begin{center}
\textbf{\textcolor{red}{low mult pp}}\\
\includegraphics[scale=0.22]{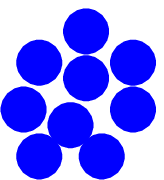}
\par\end{center}%
\end{minipage}
\par\end{centering}

\protect\caption{(Color online) Core (red) and corona (blue) in a plane perpendicular
to the beam axis, for big (left), small (right) and intermediate (middle)
systems. \label{fig-4}}
 
\end{figure}

Crucial for the following discussion is the fact that the core-corona
separation behaves quite differently for small and big systems (although
we employ the same procedure), see fig. \ref{fig-4}. 
Whereas for very large multiplicity (central AA) most of the matter
is core, with small corona contributions at the surface (or some high
pt segments further inside), we get at very low multiplicity (in pp)
only corona : the system is not dense enough to form a core. High
multiplicity pp and pA, or peripheral AA collisions are in between,
both core and corona being important.
\begin{table}[b]
\begin{tabular}{|c|c|c|c|}
\hline 
core & green dashed-dotted & particle from the core only & no hadronic cascade\tabularnewline
\hline 
corona & blue dotted & particles from corona only & no hadronic cascade\tabularnewline
\hline 
co-co & yellow dashed & particles from core and corona & no hadronic cascade\tabularnewline
\hline 
full & red full & all particles & with hadronic cascade\tabularnewline
\hline 
 & blue triangles & particles from pure string decay & no cascade, no hydro\tabularnewline
\hline 
\hline 
 & thin lines & pp simulation & \tabularnewline
\hline 
 & intermediate lines & pA (pPb) simulation  & \tabularnewline
\hline 
 & thick lines  & AA (PbPb) simulation & \tabularnewline
\hline 
\hline 
 & open circles & pp data & \tabularnewline
\hline 
 & open squares & pA (pPb) data & \tabularnewline
\hline 
 & open stars & AA (PbPb) data & \tabularnewline
\hline 
\end{tabular}\protect\caption{Color, line, and symbol codes used for the plots in this paper, to
accommodate the different contributions and the different systems.\label{tab:Color-and-line}}
\end{table}
 
\begin{figure}[tb]
\noindent \begin{centering}
~~
\par\end{centering}

\noindent \begin{centering}
(a)%
\begin{minipage}[t]{0.5\columnwidth}%
\noindent \begin{center}
\includegraphics[angle=270,scale=0.25]{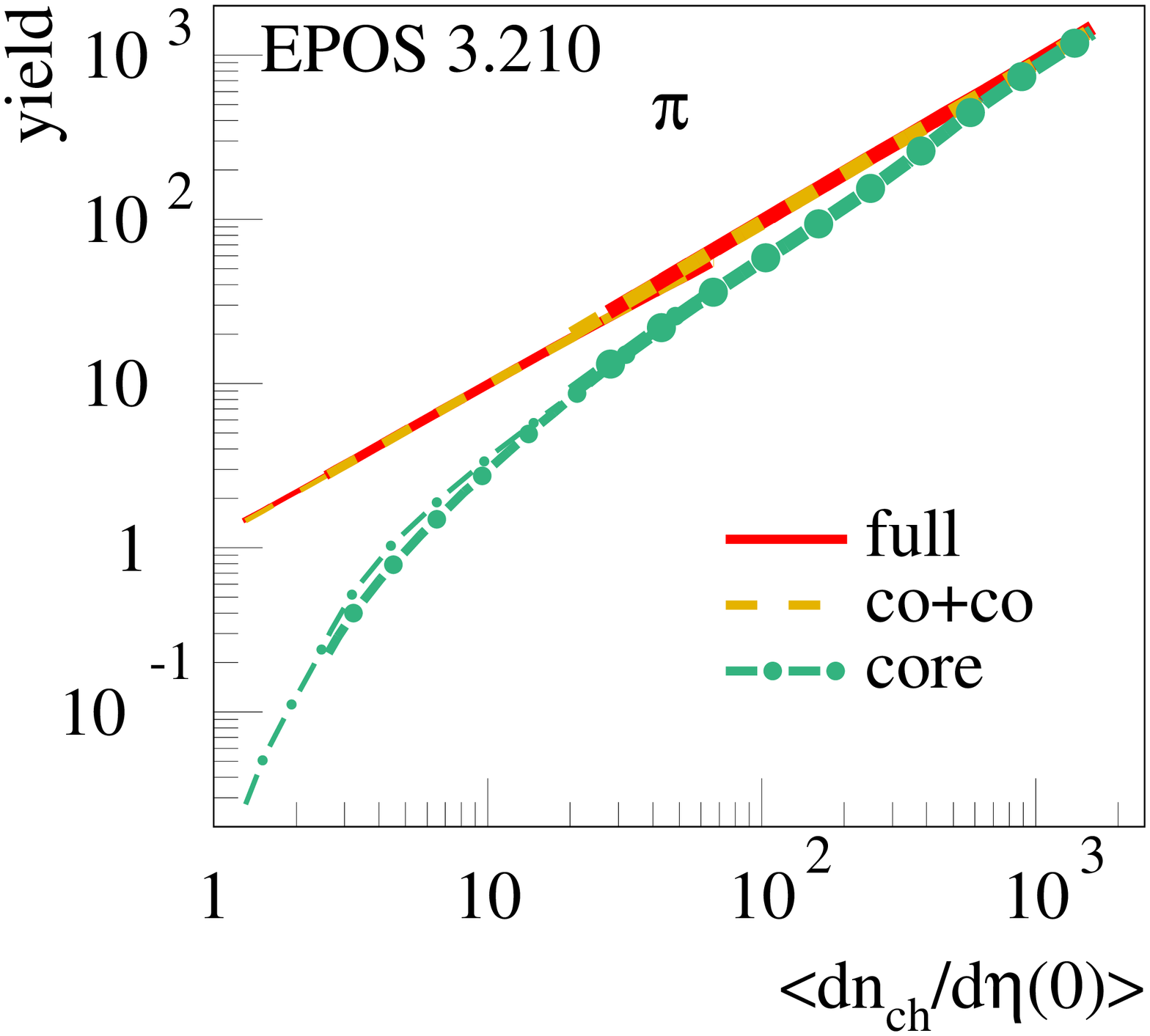}
\par\end{center}%
\end{minipage}(b)%
\begin{minipage}[t]{0.4\columnwidth}%
\noindent ~\\
~~~~~\includegraphics[angle=270,scale=0.25]{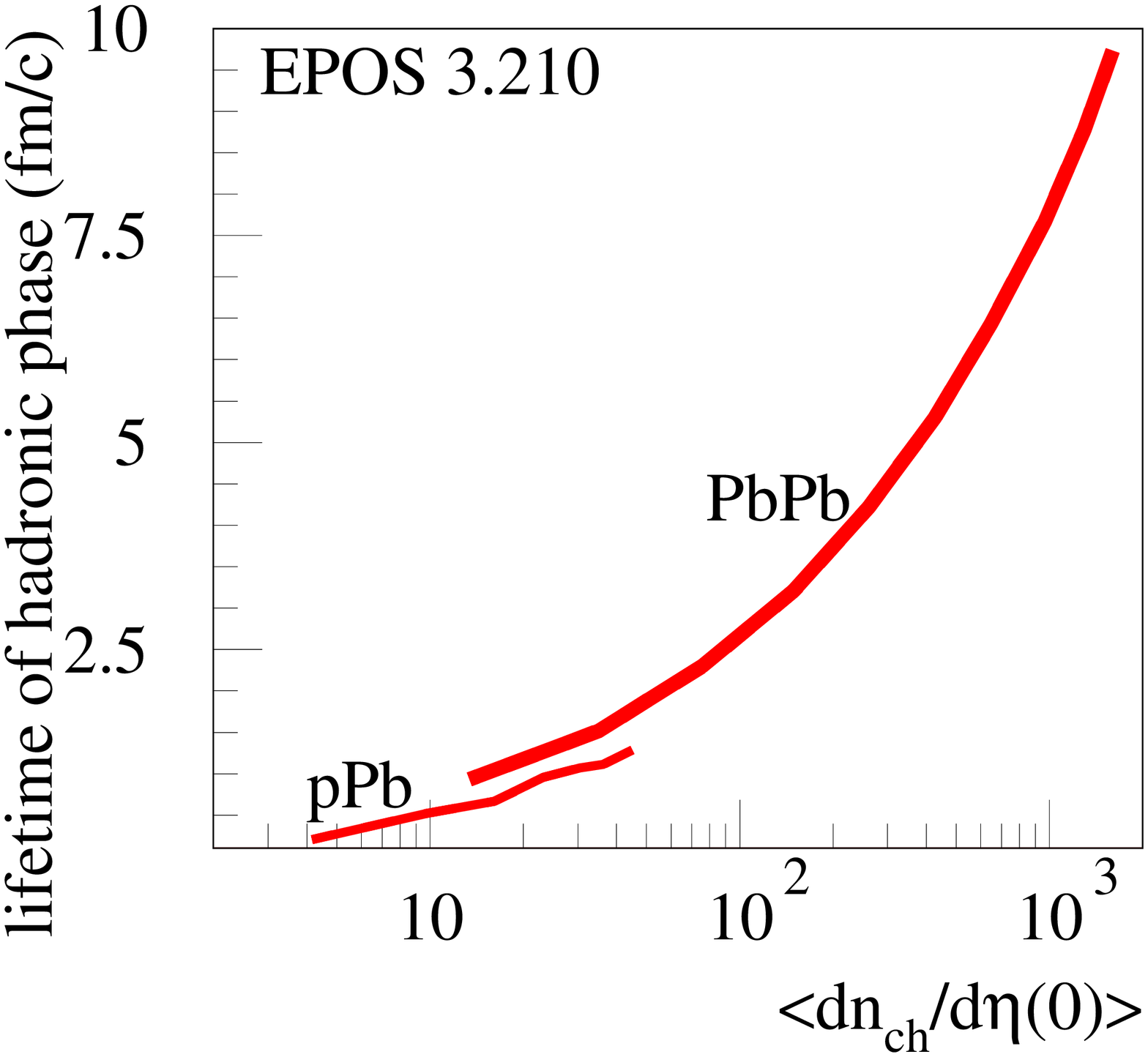}%
\end{minipage}~~~
\par\end{centering}

\protect\caption{(Color online) (a) Pion yield versus multiplicity. We compare different
contributions (core, core plus corona (co-co) and the the full contribution
(the latter on including the final state hadronic cascade) for different
systems: pp (thin lines), AA (thick lines), and pA (intermediate lines).
(A summary of color, line, and symbol codes is found in table \ref{tab:Color-and-line})
(b) Lifetimes of the hadronic phase versus multiplicity. \label{fig-6}}
 
\end{figure}
Another way of seeing the multiplicity dependence of the core-corona
separation is shown in fig. \ref{fig-6}(a), where we plot the pion
yield versus multiplicity, for only core particles (``core'', green
dashed-dotted), the contribution from core + corona (``co-co'',
particles from core and corona, yellow dashed), both contributions
from EPOS simulation without hadronic cascade. The ratio ``core''
over ``co-co'' would be the relative core fraction, which increases
continously from zero (at small multiplicity) to unity (at large multiplicity).
We also plot the ``full'' contribution, referring to all particles,
for a simulation with hadronic cascade (A summary of color, line,
and symbol codes is found in table \ref{tab:Color-and-line}). In
all cases we plot curves for the three systems: pp, pA, and AA. There
are substantial overlap regions, where different systems contribute,
but we observe unique curves, no system size dependence. We also observe
that the hadronic cascade has no effect on the pion production (``co-co''
and ``full'' gives identical results). 

In fig. \ref{fig-6}(b), we plot a quantity relevant for resonance
decays in the hadronic stage: The lifetime of this phase, defined
as the difference of the average formation time of particles for simulations
with and without hadronic cascade. The bigger the lifetime, the bigger
the probablility of a resonance decay inside this phase. The lifetime
increases continuously from less than a fm/c to 10fm/c for big systems.
\begin{figure}[tb]
\noindent \begin{centering}
(a)%
\begin{minipage}[t]{0.44\columnwidth}%
\noindent \begin{center}
\vspace{-0.25cm}
\includegraphics[angle=270,scale=0.25]{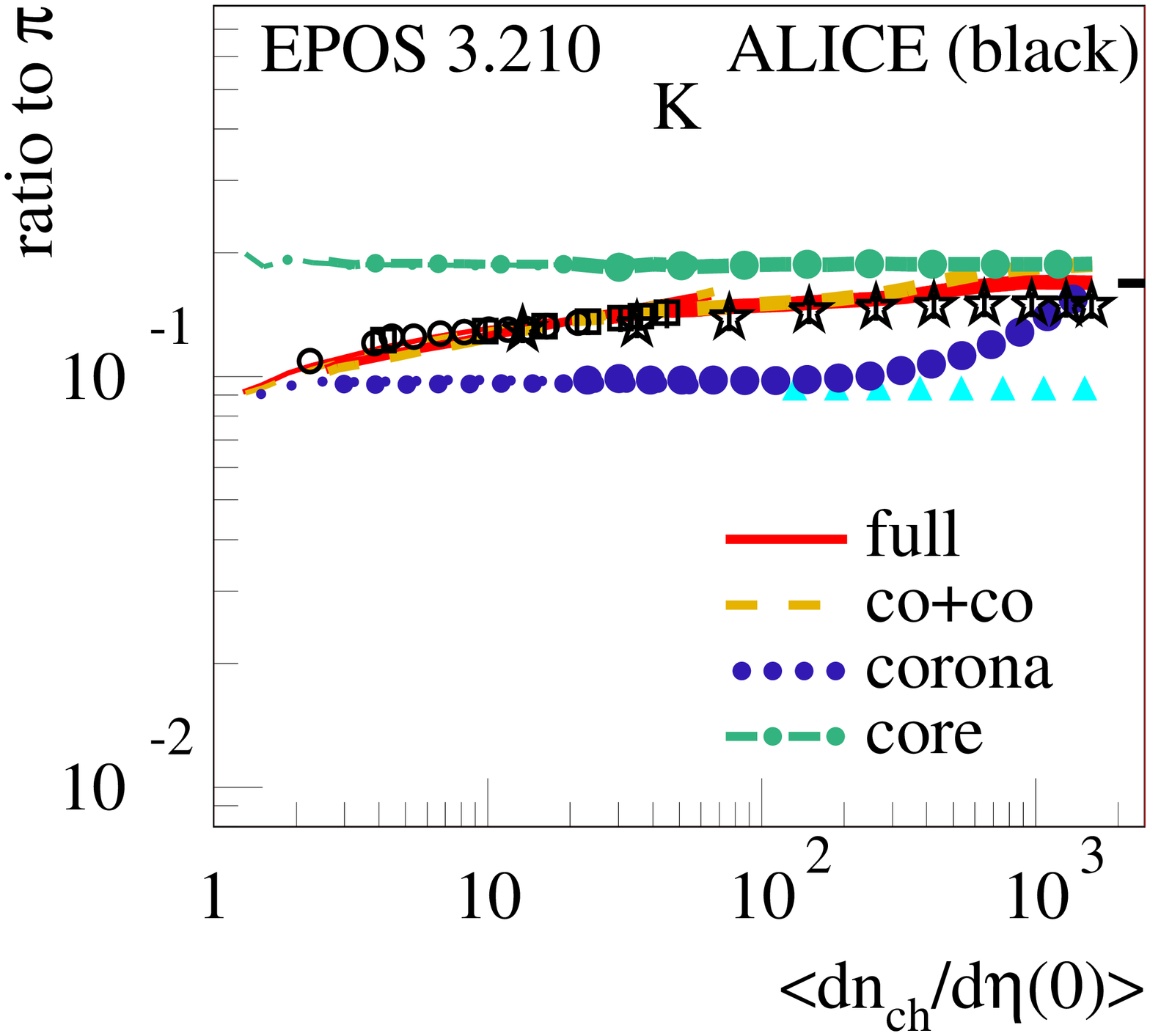}
\par\end{center}%
\end{minipage}~~~(b)%
\begin{minipage}[t]{0.44\columnwidth}%
\noindent \begin{center}
\vspace{-0.25cm}
\includegraphics[angle=270,scale=0.25]{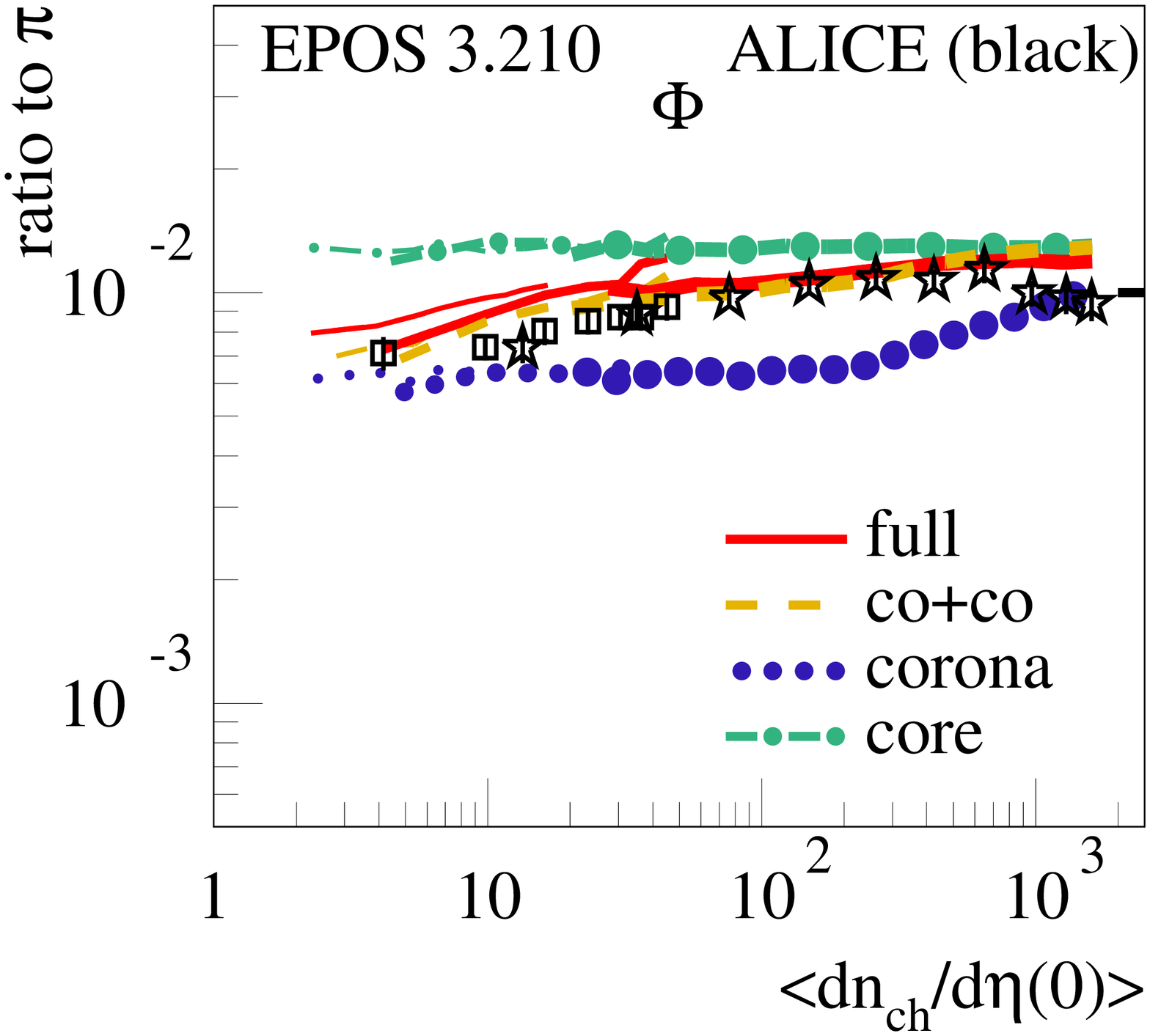}
\par\end{center}%
\end{minipage}\vspace{-0.7cm}
\\
(c)%
\begin{minipage}[t]{0.44\columnwidth}%
\noindent \begin{center}
\vspace{-0.25cm}
\includegraphics[angle=270,scale=0.25]{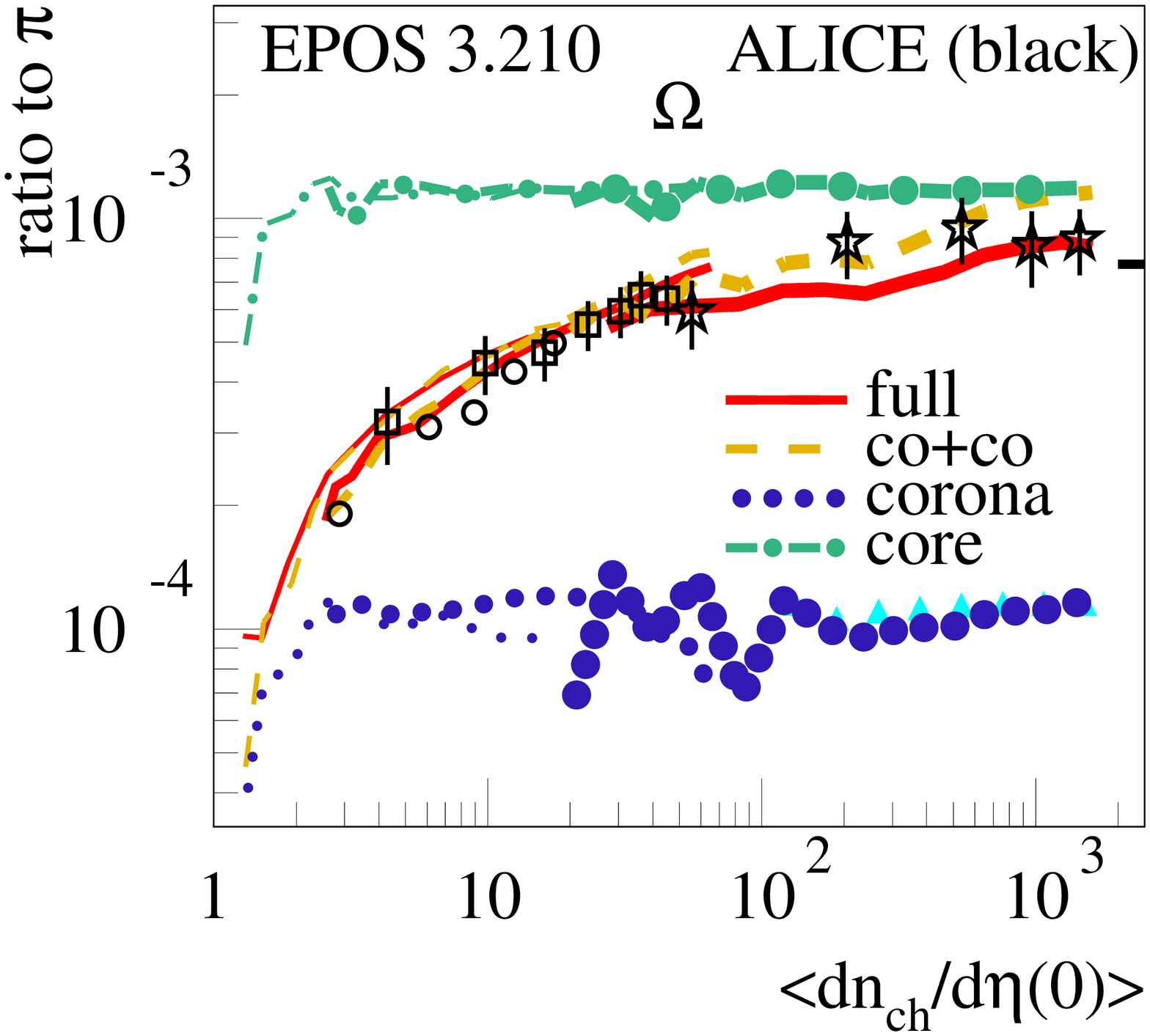}
\par\end{center}%
\end{minipage}~~~(d)%
\begin{minipage}[t]{0.44\columnwidth}%
\noindent \begin{center}
\vspace{-0.25cm}
\includegraphics[angle=270,scale=0.25]{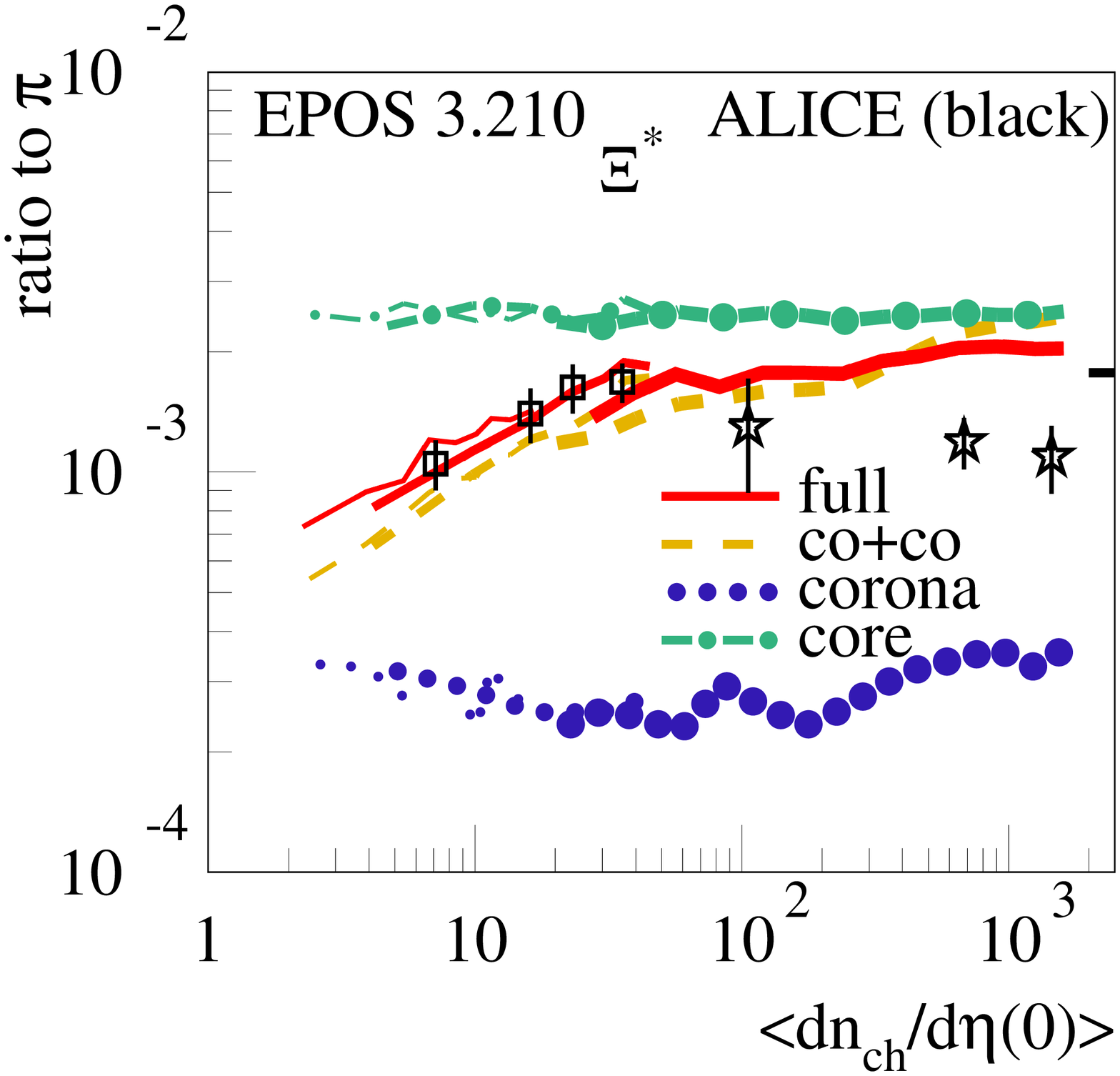}
\par\end{center}%
\end{minipage}\\
\vspace{-0.7cm}
(e)%
\begin{minipage}[t]{0.44\columnwidth}%
\noindent \begin{center}
\vspace{-0.25cm}
\includegraphics[angle=270,scale=0.25]{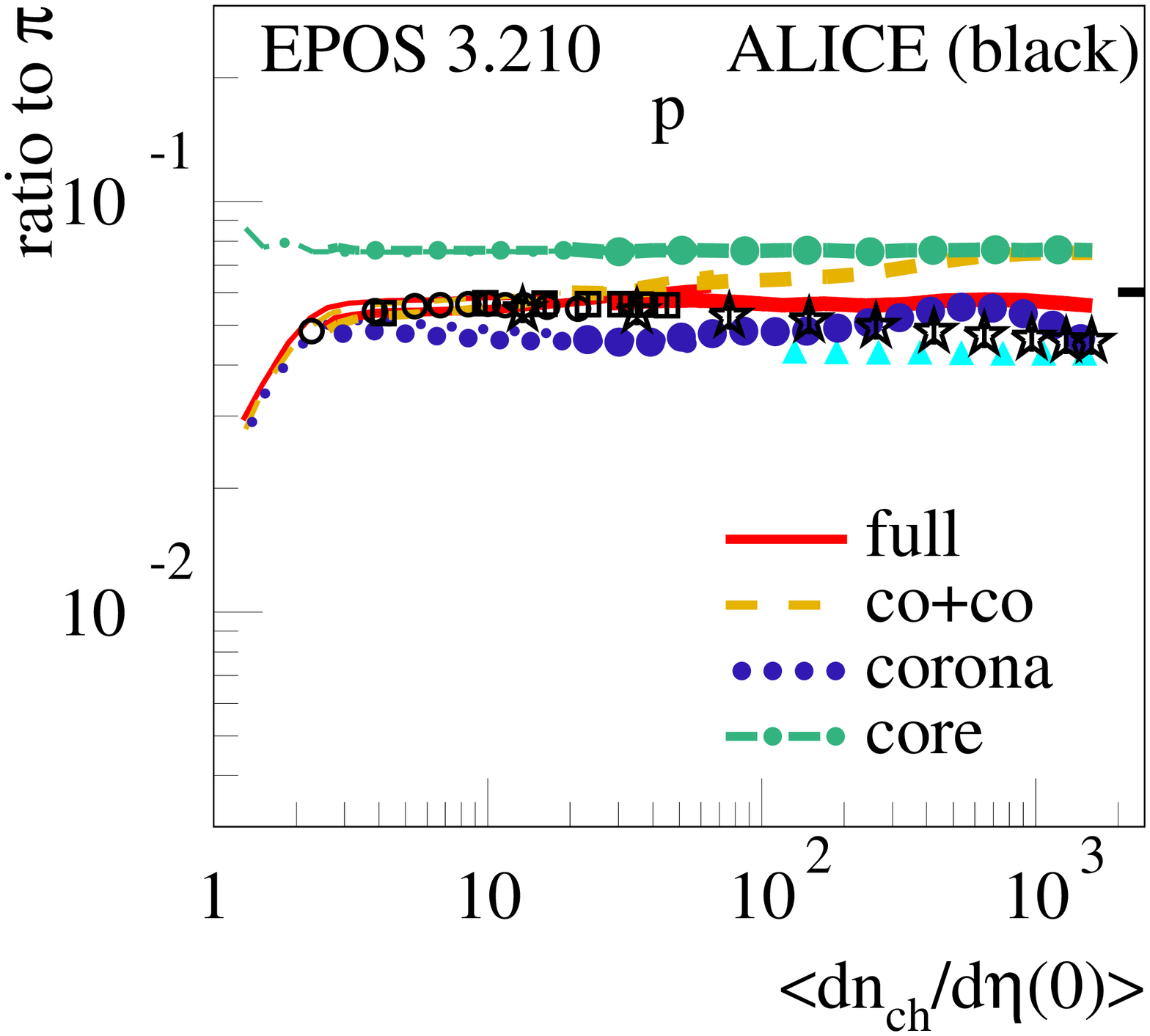}
\par\end{center}%
\end{minipage}~~~(f)%
\begin{minipage}[t]{0.44\columnwidth}%
\noindent \begin{center}
\vspace{-0.25cm}
\includegraphics[angle=270,scale=0.25]{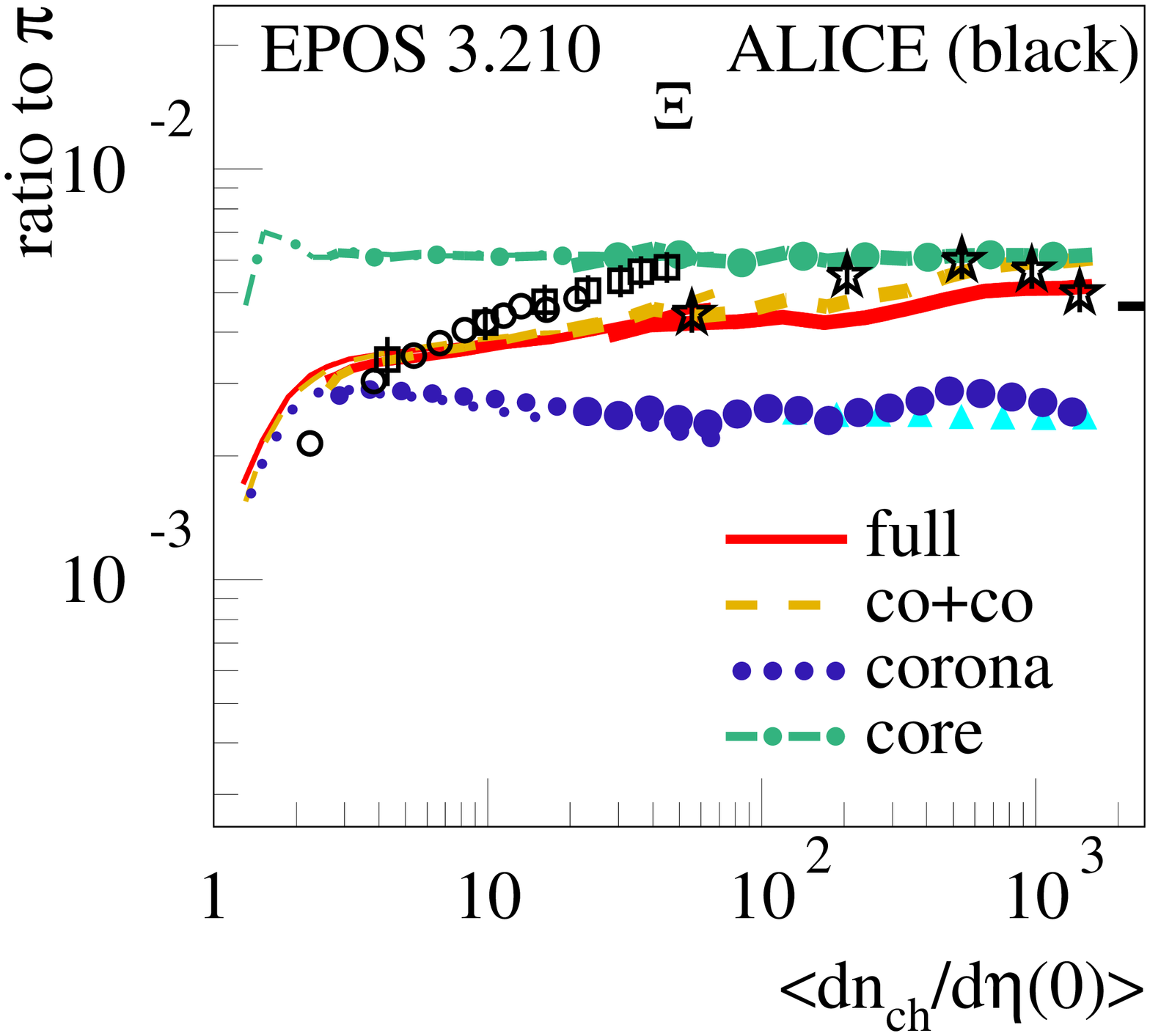}
\par\end{center}%
\end{minipage}\vspace{-0.7cm}

\par\end{centering}

\protect\caption{(Color online) Particle to pion ratio for different particle species
versus multiplicity, for different contributions from the EPOS simulations,
for different systems (pp, pA, AA). We also plot ALICE data from \cite{ali-1,ali-2,ali-3,ali-4,ali-5,ali-6,ali-7,ali-8,ali-9,ali-10,ali-11,ali-12,ali-13}.
See table \ref{tab:Color-and-line} for color, line, and symbol codes.
\label{fig-7}}
 
\end{figure}

In the following we will investigate the multiplicity dependence (for
pp, pA, AA) of particle ratios to pions, for the different contributions
from EPOS simulations, compared to data from ALICE \cite{ali-1,ali-2,ali-3,ali-4,ali-5,ali-6,ali-7,ali-8,ali-9,ali-10}.
We always refer to the color, line, and symbol codes of table \ref{tab:Color-and-line}.

In fig. \ref{fig-7}(a), we plot the \textbf{kaon} to pion ratio versus
multiplicity, for different contributions from the EPOS simulations,
for different systems (pp, pA, AA), see table \ref{tab:Color-and-line}
for color, line, and symbol codes. Despite a large overlap for the
different systems, we observe universal curves, for all contributions.
The core contribution is completely flat, as expected since the particle
ratios only depend on the properties of the fluid at freeze out, taken
to be the same in all systems. Also the corona contribution (from
string segments which escape the fluid) is essentially flat, but with
some deviation from pure string fragmentation (blue triangles). But
at large multiplicity the corona contribution is small, so we have
essentially two flat curves, from core and corona, where here in case
of kaons the core contribution is roughly a factor of 2 above the
corona. From fig. \ref{fig-6}(a), we know that the relative core
weight increases continuously from zero to unity, which explains the
increase of the core+corona (co-co) contribution which interpolates
between the corona level at small multiplicity to the core level at
high multiplicity. Considering finally the ``full'' contribution
(including the final state hadronic cascade), we see no difference
compared to the case without cascade (co-co), both curves reproducing
the experimental data. In fig. \ref{fig-7}(b), we consider $\boldsymbol{\boldsymbol{\phi}}$
\textbf{mesons}, and we observe almost the same behavior as seen for
the kaons, a continuous increase of roughly a factor of 2 from corona
to core level, and little effect of the hadronic cascade. Due to the
\textbf{long lifetime} of 46.2 fm/c only few $\phi$ mesons decay
in the hadronic phase. 

The situation is different for $\boldsymbol{\boldsymbol{\Omega}}$
\textbf{baryons}, see fig. \ref{fig-7}(c), since here core and corona
curves are separated by about a factor of 10, leading to a strong
increase of the core+corona (co-co) contribution of a factor of 10
with multiplicity. The full curve is slightly reduced at high multiplicity
compared to co-co, due to hadronic final state interactions (baryon-antibaryon
annihilation). We observe very similar results for the $\boldsymbol{\boldsymbol{\Xi^{*}}}$
\textbf{resonance} (lifetime 21.7 fm/c), but some discrepancy compared
to the data. 

Concerning \textbf{protons}, we see in fig. \ref{fig-7}(e) also first
an increase of the co-co curve, form the corona level to the core
level, and again (as for the Omega resonance) a reduction at high
multiplicity when including final state hadronic rescattering, coming
from \textbf{baryon-antibaryon annihilation}. We also plot $\boldsymbol{\boldsymbol{\Xi}}$
\textbf{baryons} in fig. \ref{fig-7}(f), showing a significant increase
with multiplicity.

\begin{figure}[tb]
\noindent \begin{centering}
(a)%
\begin{minipage}[t]{0.44\columnwidth}%
\noindent \begin{center}
\vspace{-0.25cm}
\includegraphics[angle=270,scale=0.25]{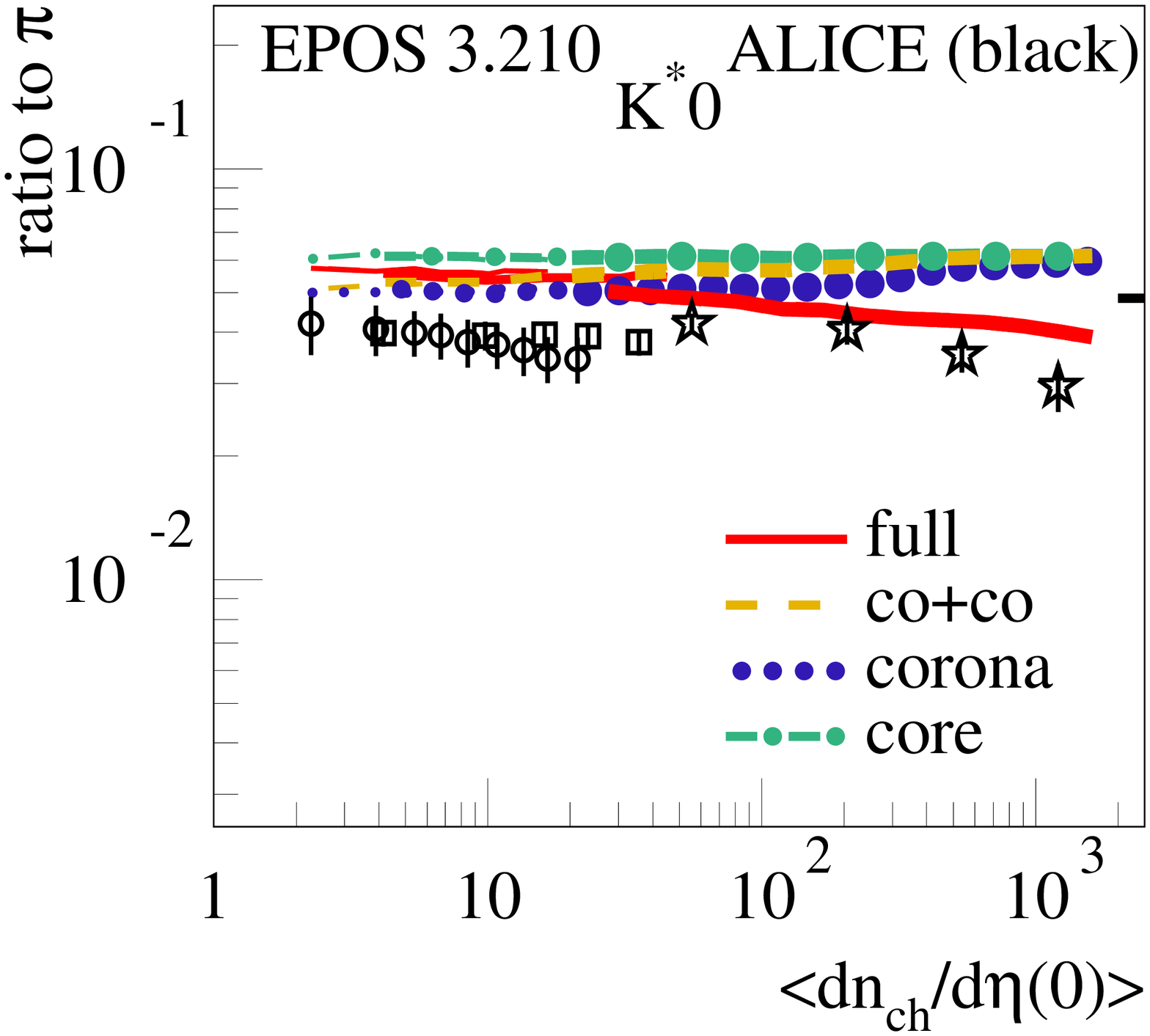}
\par\end{center}%
\end{minipage}~~~(b)%
\begin{minipage}[t]{0.44\columnwidth}%
\noindent \begin{center}
\vspace{-0.25cm}
\includegraphics[angle=270,scale=0.25]{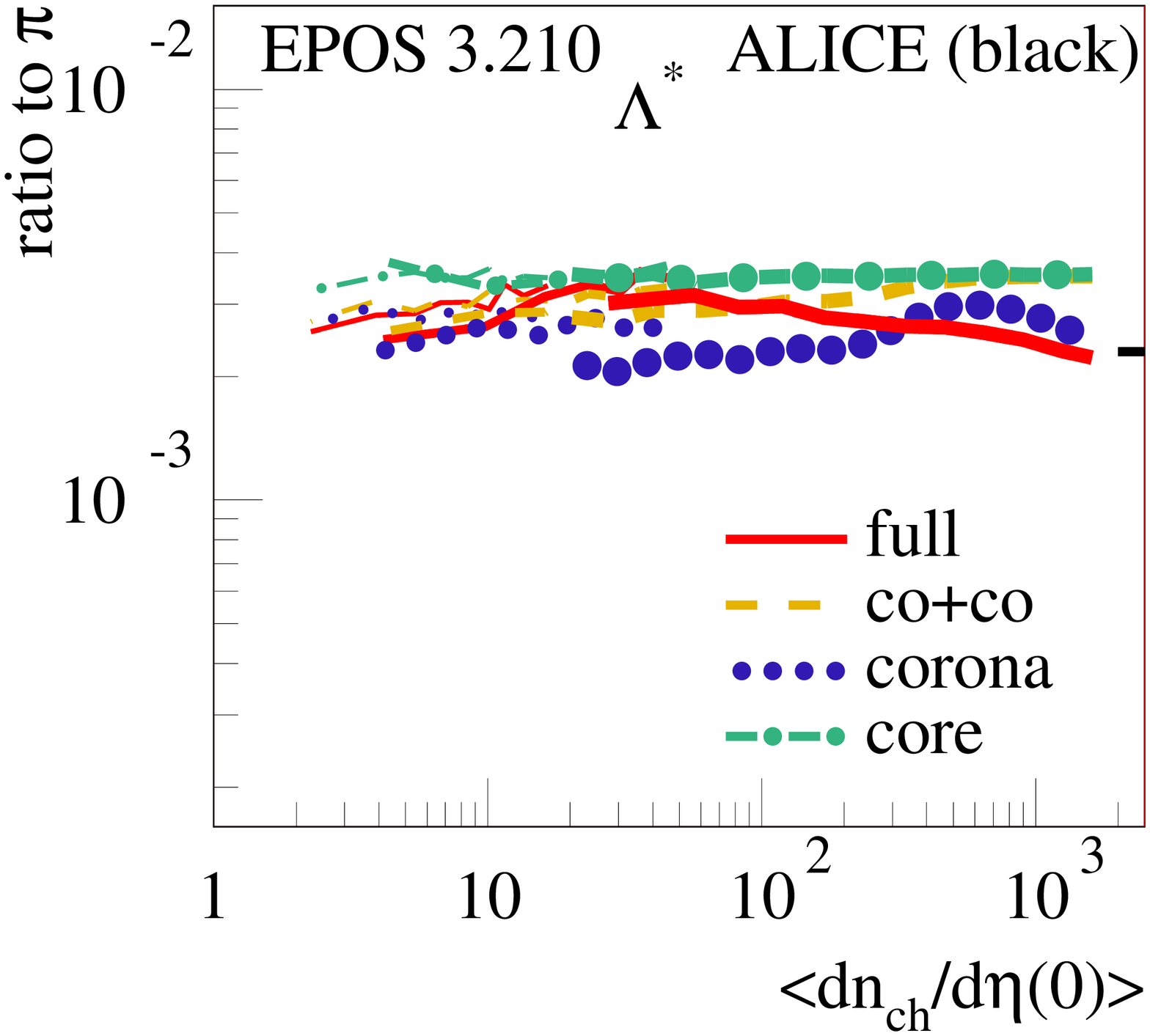}
\par\end{center}%
\end{minipage}\vspace{-0.7cm}
\\
(c)%
\begin{minipage}[t]{0.44\columnwidth}%
\noindent \begin{center}
\vspace{-0.25cm}
\includegraphics[angle=270,scale=0.25]{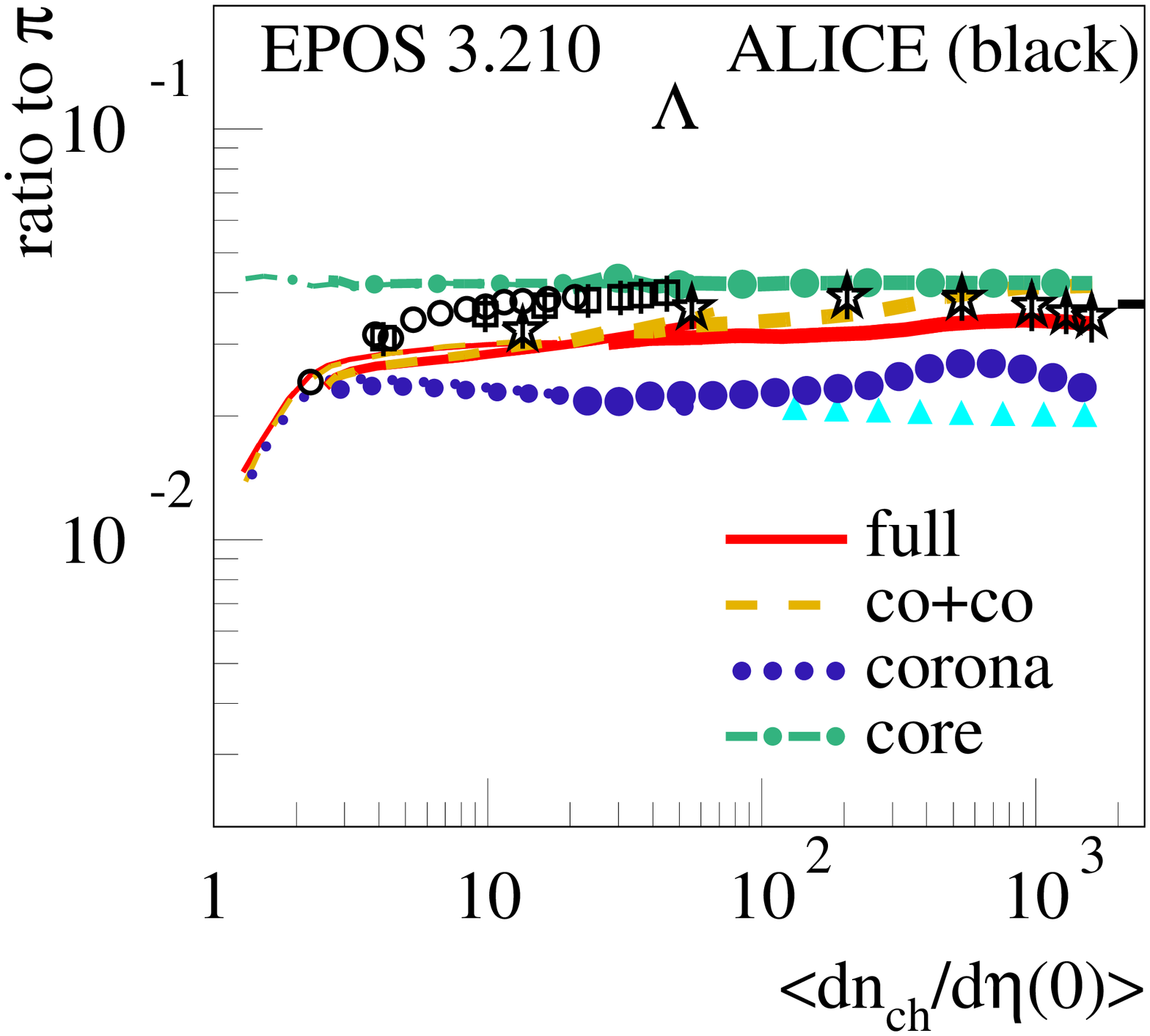}
\par\end{center}%
\end{minipage}~~~(d)%
\begin{minipage}[t]{0.44\columnwidth}%
\noindent \begin{center}
\vspace{-0.25cm}
\includegraphics[angle=270,scale=0.25]{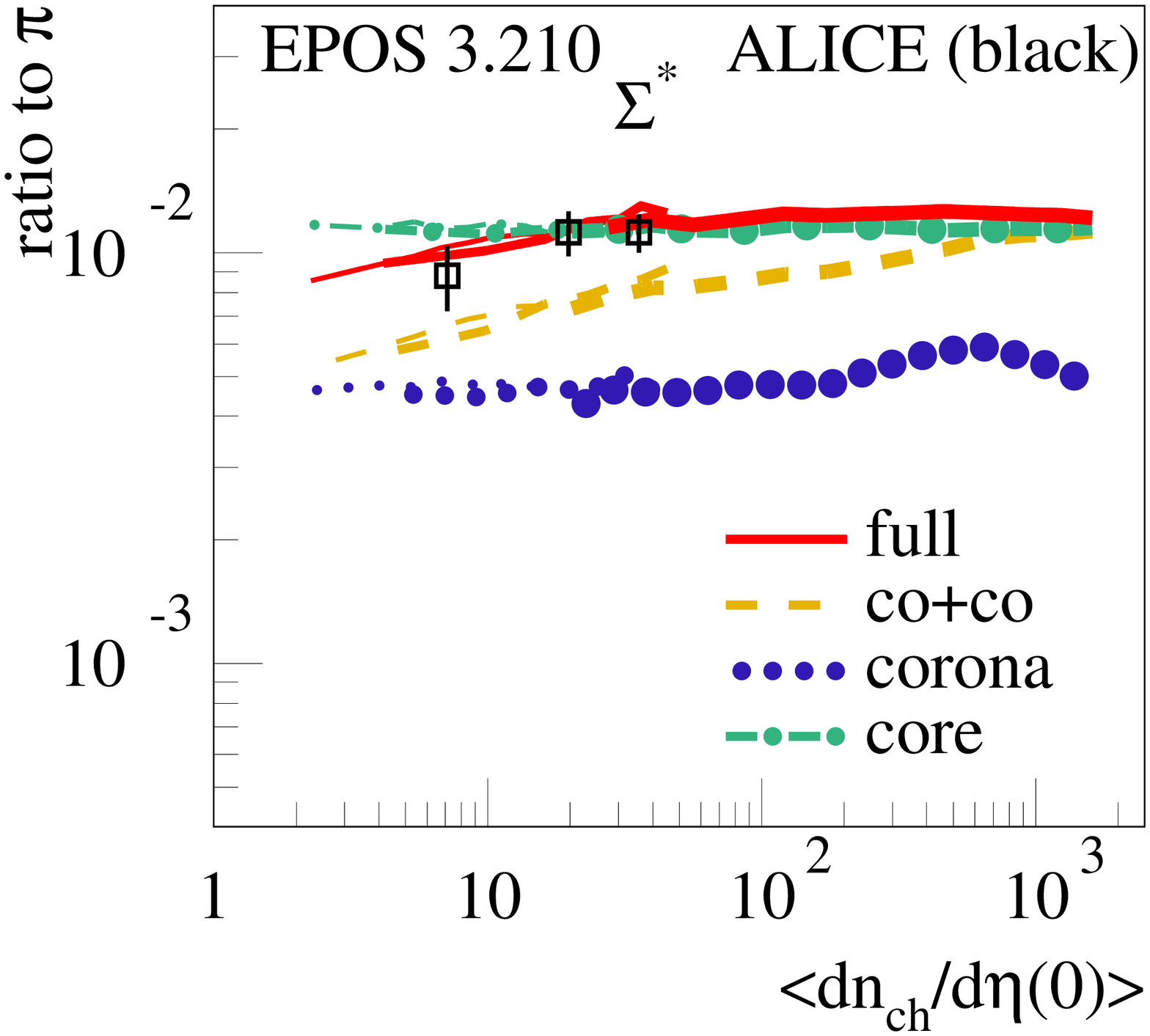}
\par\end{center}%
\end{minipage}\vspace{-0.7cm}
\\
(e)%
\begin{minipage}[t]{0.44\columnwidth}%
\noindent \begin{center}
\vspace{-0.25cm}
\includegraphics[angle=270,scale=0.25]{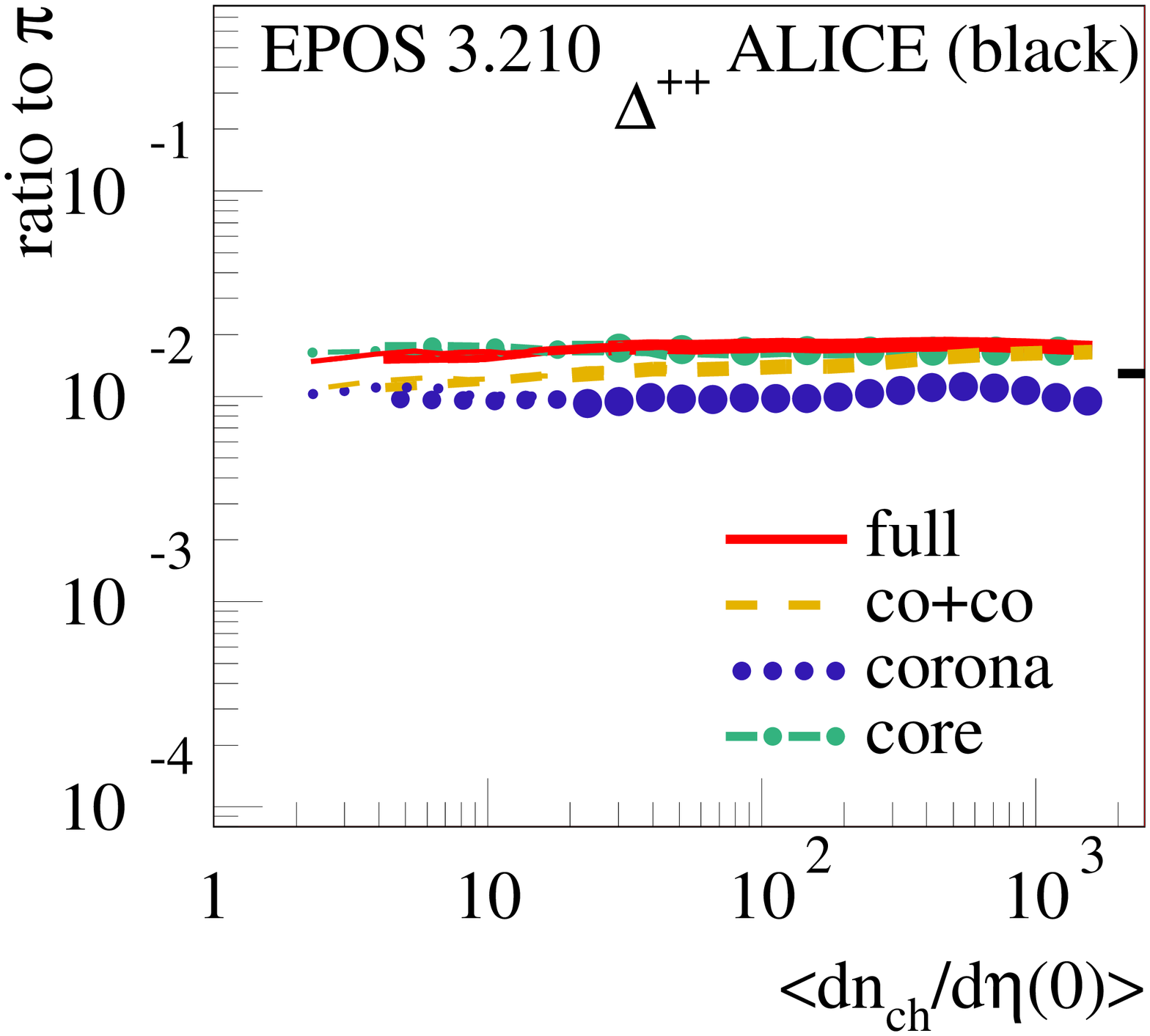}
\par\end{center}%
\end{minipage}~~~(f)%
\begin{minipage}[t]{0.44\columnwidth}%
\noindent \begin{center}
\vspace{-0.25cm}
\includegraphics[angle=270,scale=0.25]{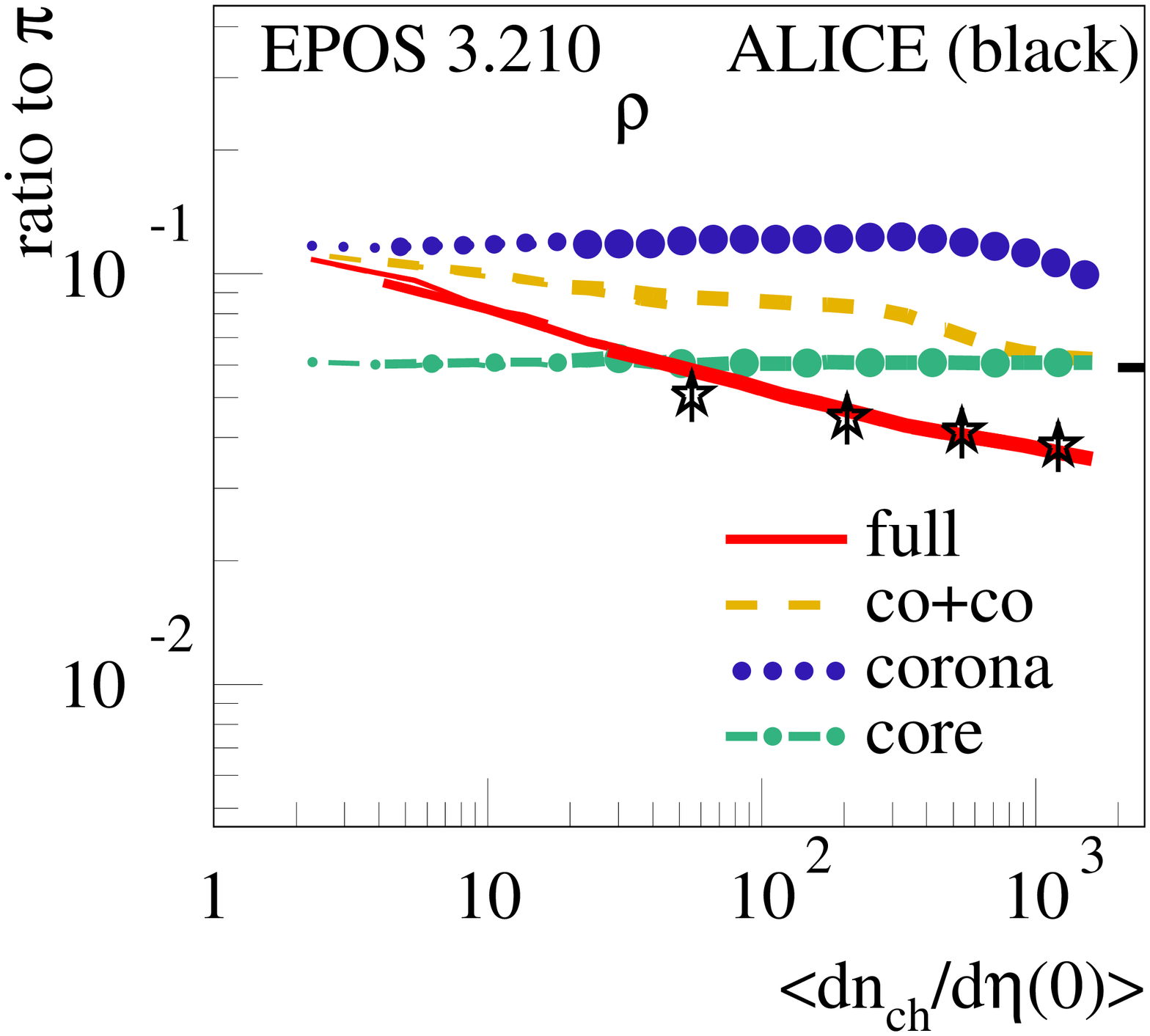}
\par\end{center}%
\end{minipage}\\
\vspace{-0.7cm}

\par\end{centering}

\protect\caption{(Color online) Same as fig \ref{fig-7}, for other particle species.
\label{fig-8}}
 
\end{figure}

In fig. \ref{fig-8}(a), we consider the \textbf{kaon resonance} $\boldsymbol{\boldsymbol{K_{0}^{*}}}$.
Here core, corona, and as a consequence also core+corona (co-co) are
identical (by accident). Interesting: The full results (including
the final state hadronic cascade) is considerably lower than the results
without cascade (co-co). This is due to the \textbf{short lifetime}
(4.2 fm/c), which makes the \textbf{particles decay within the hadronic
stage}. Some of the decay products rescatter (and change momenta)
and thus prevent the kaon resonances to be reconstructed via their
decay products. So contrary to the cases discussed earlier, we observe
here a monotonically decreasing curve, as observed in the data for
large multiplicities. The data show some discontinuity at low multiplicity,
contrary to what we observe in the simulation. Newer ALICE results
show as well a continuous behavior. Also the \textbf{Lambda resonance}
$\boldsymbol{\boldsymbol{\Lambda^{*}}}$ shows in \ref{fig-8}(b)
a significant decrease of the ``full'' contribution at high multiplicity,
partly due to decays in the hadronic stage (lifetime 12.6 fm/c) and
baryon-antibaryon annihilation. But one has to be careful not to draw
wrong conclusions. The lifetime is relatively large, so the effect
of the decay cannot be very big compared to Lambdas. On the other
hand, it seems that the corona and the core level are closer to each
other compared to the Lambda (see fig. \ref{fig-8}(c)), so the core+corona
curve increases less, and therefore the reduction due to annihilation
(present in both $\Lambda$ and $\Lambda^{*}$) provides a bigger
decrease in the case of $\Lambda^{*}$. 

The $\boldsymbol{\boldsymbol{\Sigma^{*}}}$ \textbf{resonance,} shown
in fig. \ref{fig-8}(d), shows also the ``usual'' monotonic increase
of the core+corona contribution, but here we get a strong increase
at low multiplicity but little change at high multiplicity, when considering
the ``full'' contribution including the hadronic cascade. We have
actually two competing processes: The lifetime is relatively short
(5 fm/c), so there should be some \textbf{reduction due to decay}
at high multiplicity, which is not visible because it is compensated
by an increase due to\textbf{ resonance creation }in the hadronic
stage. Whereas the reduction due to decay is only possible for sufficiently
big systems (at high multiplicity), resonance creating is possible
at any multiplicity, and is therefore visible at small multiplicities.
The $\boldsymbol{\boldsymbol{\Delta^{++}}}$ \textbf{resonance} (lifetime
1.7 fm/c), considered in fig. \ref{fig-8}(e), shows a similar behavior,
on a smaller scale. 

An exceptional case is the $\boldsymbol{\boldsymbol{\rho}}$ \textbf{resonance},
shown in fig. \ref{fig-8}(f), since here the core level is below
the corona one (this is the only case, it happens because $\rho$
production is very frequent in the string breaking procedure), and
therefore the \textbf{core+corona contribution decreases} with multiplicity.
Considering the full contribution (including the hadronic cascade),
there is an additional reduction, such that altogether there is a
very strong decrease of the $\rho$ over pion yield with multiplicity. 

To summarize, we have shown that the study of the multiplicity dependence
of particle ratios for different collision systems (pp, pA, AA) provides
a wealth of information concerning the production mecanism of particles
and about the properties of the medium having produced or affected
these particles.

\subsubsection*{Acknowlegments}

This research was carried out within the scope of the GDRE (European
Research Group) ``Heavy ions at ultrarelativistic energies'', and
was partially supported by COST Action CA15213 (THOR). This work was
supported by U.S. Department of Energy Office of Science under contract
number DE-SC0013391. The authors acknowledge the Texas Advanced Computing
Center (TACC) at the University of Texas at Austin for providing computing
resources that have contributed to the research results reported within
this paper. URL: http://www.tacc.utexas.edu. B. G. gratefully acknowledges
generous support from Chilean FONDECYT grants 3160493 and Proyecto
Basal FB0821. 
%%L. McLerran, R. Venugopalan, Phys. Rev. D 49 (1994) 2233; L. McLerran, R. Venugopalan, Phys. Rev. D 49 (1994) 3352; L. McLerran, R. Venugopalan, Phys. Rev. D 50 (1994) 2225.
%%pp references:
%%Alice collaboration, Phys. Lett. B 758 389-401 (2016)
%%Alice collaboration, Eur. Phys. J. C 68 345-354 (2010)
%%Alice collaboration, Eur. Phys. J. C 75 226 (2015)
%%
%%ali-7: $ \Xi  \Omega $ in Pb+Pb
%%ali-8: $ K_{0}^{*}/\pi \phi/\pi  $ in Pb+Pb
%%ali-9: data tables: https://www.hepdata.net/record/77284 $ K_{s}^{0}/\pi p/\pi \Lambda/\pi  \Xi/\pi  \Omega/\pi $  in p+p at 7 TeV
%%ali-10: $ K/\pi $ in p+p at 7 TeV 
%%ali-11: $ \Xi_{0}^{*}/\pi $ in Pb+Pb
%%ali-12: $  \rho^{0}/\pi $ in Pb+Pb
%%ali-13: $ \Sigma_{\pm}^{*}/\pi  \Xi_{0}^{*}/\pi $  in p+Pb

\end{document}